\documentclass[journal]{IEEEtran}

\usepackage[normalem]{ulem}

\usepackage{amsmath,amsfonts,amsxtra,amssymb,latexsym,amscd,amsthm,mathrsfs,bm}
\usepackage{graphicx}
\usepackage[belowskip=0pt,aboveskip=0pt,small]{caption}
\usepackage{subcaption}
\usepackage[linesnumbered, algo2e, ruled,norelsize]{algorithm2e}
\SetArgSty{textnormal}
\usepackage{threeparttable}
\usepackage{threeparttablex}
\usepackage{slashbox}
\usepackage{verbatim}
\usepackage{subfloat}
\usepackage{tabu}
\usepackage{booktabs}
\usepackage{float}
\usepackage{url}
\usepackage{multirow}
\usepackage{cite}
\usepackage{footnote}
\usepackage{array}
\usepackage[table]{xcolor}
\usepackage{tabularx}

\usepackage{algpseudocode}
\usepackage{algorithm}

\newcolumntype{P}[1]{>{\centering\arraybackslash}p{#1}}

\usepackage{epstopdf}					
\usepackage{amsthm} 
\theoremstyle{theorem}
\allowdisplaybreaks
\DeclareMathOperator*{\argmax}{argmax} 
\def\BibTeX{{\rm B\kern-.05em{\sc i\kern-.025em b}\kern-.08em
    T\kern-.1667em\lower.7ex\hbox{E}\kern-.125emX}}

\title{Multi-level Design for Multiple-Symbol Non-Coherent Unitary Constellations for Massive SIMO Systems
\thanks{The authors are with the Department of Electrical and Computer Engineering, University of Saskatchewan, Saskatoon, Canada S7N 5A9. Emails: \{wve765, ha.nguyen, e.bedeer\}@usask.ca.}
\thanks{The co-authors dedicate this work to Prof. Ha H. Nguyen, who passed away before the submission of the paper.}
}
\author{Son T. Duong, Ha H. Nguyen, and Ebrahim Bedeer}

\IEEEaftertitletext{\vspace{-2.0\baselineskip}}
\begin{document}

\maketitle

\begin{abstract}
This paper investigates non-coherent detection of single-input multiple-output (SIMO) systems over block Rayleigh fading channels. Using the Kullback-Leibler divergence as the design criterion, we formulate a multiple-symbol constellation optimization problem, which turns out to have high computational complexity to construct and detect. We exploit the structure of the formulated problem and decouple it into a unitary constellation design and a multi-level design. The proposed multi-level design has low complexity in both construction and detection. Simulation results show that our multi-level design has better performance than traditional pilot-based schemes and other existing low-complexity multi-level designs.

\end{abstract}

\begin{IEEEkeywords}
Constellation design, energy encoding and decoding, Kullback-Leibler (KL) divergence, non-coherent detection, ultra-reliable low-latency communications (URLLC).
\end{IEEEkeywords}

\section{Introduction}

Ultra-reliable low-latency communication (URLLC) has been introduced as one of the pillars of the fifth generation (5G) and beyond cellular networks. The critical requirements of URLLC, i.e., block error rate (BLER) $< 10^{-5}$ and latency $<$ 1 ms, are expected to facilitate emerging applications such as self-driving cars, and industrial automation \cite{chen2018ultra}.
Short packet communications (SPCs) with packets length of tens of bytes have been widely adopted to achieve low latency, while massive multiple-antenna systems with coherent detection have been well-studied as a diversity resource to achieve the stringent BLER requirement. 
However, to enable coherent detection, the transmitter must acquire accurate channel state information by sending a long sequence of training symbols, which decreases the spectral efficiency (SE) of SPCs. Thus, it is important to investigate the non-coherent detection of SPCs.

Among different non-coherent detection schemes, energy-based transmission and detection have been well-studied in massive single-input multiple-output (SIMO) systems \cite{chowdhury2016scaling, han2022constellation}. However, non-coherent energy-based schemes suffer from loss of the SE since only the amplitude of the signal carries information, and only one symbol is encoded and decoded at a time. That said, and to improve the SE, we aim to design non-coherent constellations that carry information in both amplitude and phase and are encoded and decoded over several symbols duration.
Please note that non-coherent detection of information in the phase is possible at the receiver by calculating the phase difference between symbols.

For designing non-coherent constellations over several symbols, we utilize the Kullback-Leibler (KL) divergence, which is recognized as an effective design criterion for non-coherent communications \cite{borran2003design}. In particular, the authors in \cite{borran2003design} proved that the pairwise error probability performance achieved by the maximum likelihood (ML) detector of any two symbols is bounded by its KL distance. This means that by maximizing the KL divergence between any two symbols, one also maximizes the symbol error rate (SER) performance of the constellation. By utilizing the KL divergence as a constellation design criterion, the authors in \cite{borran2003design} divided the constellation into different energy levels, where each level contains constellation points with the same energy. The optimization of the energy levels is called multi-level design while designing the constellation points within the same energy level is called unitary constellation design.

While researchers had focused on designing unitary constellations, e.g., see \cite{gohary2009noncoherent,zhang2011full,ngo2019cube} and the references therein, the multi-level design over several symbols has attracted little attention \cite{borran2003design,li2021constellation}. 
For instance, the authors in \cite{borran2003design} found the optimal multi-level design using an exhaustive search, and the ML detection of the optimal multi-level design was also obtained through an exhaustive search. The prohibitive computational complexity of the exhaustive search for the construction and the detection of the multi-level design in \cite{borran2003design} limits its application to a small number of symbols.
 
In \cite{li2021constellation}, the authors provided a low-complexity multi-level design over only two symbols. However, the design in \cite{li2021constellation} applies only to a specific unitary constellation and cannot be generalized to other structures or to other numbers of symbols.

To the best of our knowledge, a low-complexity multi-level design over any number of symbols for non-coherent detection of SPCs has not been considered in the literature. That said, we design a multi-level scheme based on the KL divergence to improve the SER performance of existing non-coherent unitary constellations designs. The proposed multi-level constellation has low complexity in both its construction and detection to meet the latency requirement of URLLC systems. Simulation results show that our multi-level design significantly improves the SER performance of existing unitary constellations and outperforms other low-complexity non-coherent schemes.

\emph{Notation:} Matrices, column vectors, and scalar variables are denoted by uppercase bold letters (e.g., $\mathbf{A}$), lowercase bold letters (e.g., $\mathbf{a}$) and lowercase letters (e.g., $a$), respectively. We use $(\cdot)^*$, $(\cdot)^T$, and $(\cdot)^H$ to denote the conjugate, transpose, and conjugate transpose, respectively. We use $|\cdot|$, $\Vert \cdot \Vert$, $\det(\cdot)$, and $\rm{tr}(\cdot)$ to denote absolute value, Euclidean norm, determinant, and trace operations, respectively. We use $\mathbb{R}^{m \times n}$ and $\mathbb{C}^{m \times n}$ to indicate the set of real and complex matrices with dimension $m \times n$, respectively. $\mathbf{I}_K$ denotes the $K \times K$ identity matrix and $\rm{card}\{\Omega\}$ represents the cardinality of the set $\Omega$. The circularly symmetric complex Gaussian distribution with mean $\mu$ and variance $\sigma^2$ is represented as $\mathcal{CN}(\mu,\sigma^2)$, and we use $\mathbb{E}(\cdot)$ to denote the expectation of a random variable.

The remainder of the paper is organized as follows. The system model is presented in Section~\ref{section:system_model}. We overview the design method based on KL divergence and propose our multi-level structure in Section~\ref{section:KL-based_design}. The optimization formulation of our proposed multi-level structure is provided in Section~\ref{section:optimization}. Simulation results and related discussion are given in Section~\ref{section:simulation_result}. Section~\ref{section:conclusion} concludes the paper.

\section{System Model}
\label{section:system_model}
Consider the uplink transmission of a SIMO system in which a single-antenna transmitter communicates with an $M$-antenna base station. Let $\mathbf{h} \sim \mathcal{C N}(0, \mathbf{I}_M)$ be the vector of independent Rayleigh fading channel coefficients between the transmitter and the receiver, and it is unknown to both of them. We assume a block fading channel where the channel coefficients stay constant over a block of consecutive $K$ symbols and then change to an independent realization in the coming block. Let $\mathbf{s} \in \mathbb{C}^{K \times 1}$ be a sequence of $K$ transmit symbols that represents one constellation point. The received signal $\mathbf{Y} \in \mathbb{C}^{M \times K}$ at the receiver can be formulated as:
\begin{equation}
    \mathbf{Y} = \mathbf{h} \mathbf{s}^T + \mathbf{N},
\end{equation}
where $\mathbf{N} \in \mathbb{C}^{M \times K}$ is the noise matrix, whose elements are independent zero-mean circular Gaussian random variables with variance $\sigma^2$. The transmitted sequence is assumed to satisfy the power constraint $\mathbb{E}(\Vert \mathbf{s} \Vert^2) = 1$.
Thus, the average SNR at each receiving antenna is given as $\text{SNR}=1/(K\sigma^2)$.

Given the transmit signal $\mathbf{s}$, the probability density function of the received signal $\mathbf{Y}$  can be written as follows \cite{borran2003design}, \cite{li2021constellation}:
\begin{align}
    f(\mathbf{Y}|\mathbf{s}) & = \frac{\exp\left(-\rm{tr}\left(\mathbf{Y}^H\mathbf{Y}(\mathbf{s}^*\mathbf{s}^T+\sigma^2\mathbf{I}_K)^{-1}\right)\right)}{\pi^{KM}\left[\det(\mathbf{s}^*\mathbf{s}^T + \sigma^2 \mathbf{I}_K)\right]^{M}}, \nonumber \\
    & = \frac{\exp\left(-\frac{\rm{tr}(\mathbf{Y}^H\mathbf{Y})}{\sigma^2} + \frac{\rm{tr}\left( \mathbf{Y}^H \mathbf{Y} \ \mathbf{s}^* \mathbf{s}^T \right)}{\sigma^2(\sigma^2 + \Vert \mathbf{s} \Vert^2)} \right)}{\pi^{KM}\left[ (\sigma^2 + \Vert \mathbf{s} \Vert^2)\sigma^{2K-2} \right]^{M}}.
    \label{eqn:density_func}
\end{align}
Given a known constellation of $\mathbf{s}$, i.e., $\Omega_s$, the ML detector to detect the transmit signal $\mathbf{s}$ is \cite{li2021constellation}:
\begin{equation}
    \mathbf{s}_{\rm{ML}} = \argmax_{\mathbf{s} \in \Omega_s} \left( \frac{\rm{tr} \left( \mathbf{Y}^H \mathbf{Y} \ \mathbf{s}^* \mathbf{s}^T \right)}{\sigma^2(\sigma^2 + \Vert \mathbf{s} \Vert^2 )} - M (\sigma^2 + \Vert \mathbf{s} \Vert^2 ) \right).
    \label{eqn:ML_detector}
\end{equation}

For any two transmit signals $\mathbf{s}_i$ and $\mathbf{s}_k$, the KL distance between them is defined as the average KL divergence of their two conditional probability density functions per antenna, 
which can be formally written as\cite{borran2003design}:
\begin{align}
    \label{eq:KL_def}
    & D_{\rm{KL}} \left( f(\mathbf{Y}|\mathbf{s}_i), f(\mathbf{Y}|\mathbf{s}_k) \right) = \frac{1}{M} \mathbb{E}_{f(\mathbf{Y}|\mathbf{s}_i)} \left( \frac{f(\mathbf{Y}|\mathbf{s}_i)}{f(\mathbf{Y}|\mathbf{s}_k)} \right) \nonumber \\
    & = \left( \frac{ \Vert \mathbf{s}_k \Vert ^2  \Vert \mathbf{s}_i \Vert ^2 - \Vert \mathbf{s}_k^T \mathbf{s}_i^* \Vert ^2}{\sigma^2 (\sigma^2 +  \Vert \mathbf{s}_k \Vert ^2)} \right) \nonumber \\
    & \qquad + \left( \frac{\sigma^2 +  \Vert \mathbf{s}_i \Vert ^2}{\sigma^2 +  \Vert \mathbf{s}_k \Vert ^2} - \rm{ln} (\frac{\sigma^2 +  \Vert \mathbf{s}_i \Vert ^2}{\sigma^2 +  \Vert \mathbf{s}_k \Vert ^2}) - 1 \right).
\end{align}

\section{KL-Based Multi-Level Constellation Design}
\label{section:KL-based_design}
As pointed out in \cite{borran2003design}, the KL distance between any two transmit sequences $\mathbf{s}_i$ and $\mathbf{s}_k$ determines the pairwise symbol error probability between them. For this reason, the pair with the smallest KL distance will produce the highest error probability and contribute the most to the error rate of the whole constellation set. As a result, we aim to design a constellation set $\Omega_s$ that maximizes the minimum KL distance, which can be formally expressed as follows:
\begin{align}
    \max_{\Omega_s} \min_{\mathbf{s}_i,\mathbf{s}_k \in \Omega_s, \mathbf{s}_i \neq \mathbf{s}_k} D_{\rm{KL}} \left( \mathbf{s}_i, \mathbf{s}_k \right). \label{eq:main}
\end{align}

Since any sequence $\mathbf{s}$ is a $(K\times1)$-dimensional complex vector, it can be characterized by its magnitude $\alpha \in \mathbb{R}$ and its direction vector $\mathbf{v} \in \mathbf{C}^{K \times 1}$, which are defined as follows:
\begin{align}
    \alpha = \Vert \mathbf{s} \Vert , \ \mathbf{v} = \frac{\mathbf{s}}{\Vert \mathbf{s} \Vert}.
\end{align}
Mathematically, $\mathbf{v}$ is a unitary vector obtained by normalizing the original vector $\mathbf{s}$ such that any two sequences $\mathbf{s}_i$ and $\mathbf{s}_k$ with the same direction will have the same unitary vector $\mathbf{v}$.  

The maximum likelihood detector in (\ref{eqn:ML_detector}) can be rewritten in terms of $\alpha$ and $\mathbf{v}$  as follows:
\begin{equation}
    \{ \alpha_{\rm{ML}}, \mathbf{v}_{\rm{ML}} \} = \argmax_{(\alpha,\mathbf{v})} \left( \frac{\alpha^2\rm{tr} \left( \mathbf{Y}^H \mathbf{Y} \mathbf{v}^* \mathbf{v}^T \right)}{\sigma^2(\sigma^2 + \alpha^2 )} - M (\sigma^2 + \alpha^2 ) \right).
    \label{eqn:ML_detector_splitted}
\end{equation}
Similarly, the KL divergence in \eqref{eq:KL_def} is also rewritten in terms of $\alpha$ and $\mathbf{v}$  as follow:
\begin{align}
    & D_{\rm{KL}} \left( \alpha_k,\mathbf{v}_k,\alpha_i,\mathbf{v}_i \right) = D_{1}\left( \alpha_k,\mathbf{v}_k,\alpha_i,\mathbf{v}_i \right) + D_{2}\left( \alpha_k,\alpha_i \right), \nonumber \\
    & = \frac{ \alpha_k^2 \alpha_i^2 \left( 1 - | \mathbf{v}_k^T \mathbf{v}_i^* | ^2 \right) }{\sigma^2 (\sigma^2 + \alpha_k^2)} + \left( \frac{\sigma^2 + \alpha_i^2}{\sigma^2 + \alpha_k^2} - \rm{ln} \left(\frac{\sigma^2 + \alpha_i ^2}{\sigma^2 + \alpha_k^2}\right) - 1 \right).
    \label{eq:KL_divergence_after_transformation}
\end{align}
As can be seen from \eqref{eq:KL_divergence_after_transformation}, the KL distance between two constellation points $\mathbf{s}_i$ and $\mathbf{s}_k$ consist of two terms $D_1$ and $D_2$. The first term $D_1$ is due to the difference of direction vectors $\mathbf{v}_i$ and $\mathbf{v}_k$, and it is scaled by the energy levels $\alpha_i^2$ and $\alpha_k^2$ of the two points. On the other hand, the second term $D_2$ is mainly due to the difference between the energy levels of the two points.

Equation \eqref{eq:KL_divergence_after_transformation} indicates that designing the optimal $\Omega_{s}$ based on the KL divergence requires a joint optimization over both $\alpha$ and $\mathbf{v}$, which is an extremely complex process. One can observe that the KL divergence in (\ref{eq:KL_divergence_after_transformation}) reduces to $D_2$, i.e., $D_1=0$, when the two constellation points have the same direction $\mathbf{v}$ but different energies; and reduces to $D_1$, i.e., $D_2=0$, when the two constellation points have the same energy levels. That said, to strike a balance between the SER performance and complexity, we split the constellation set $\Omega_s$ into different subsets $\mathcal{W}_n$, $n = 0, ..., N-1$, 
where each subset $\mathcal{W}_{n}$ comprises of $w_n$ points with the same energy level $\alpha_n^2$. By doing so, we transform the constellation design problem in \eqref{eq:KL_divergence_after_transformation} into optimizing the KL distance between constellation points inside a given subset $D_{\rm{intra}}(\mathcal{W}_n)$ (which only contains $D_1$) and optimizing the KL distance between different subsets $D_{\rm{inter}}(\mathcal{W}_{n_1},\mathcal{W}_{n_2})$. Please note that both $D_{\rm{intra}}(\mathcal{W}_n)$ and $D_{\rm{inter}}(\mathcal{W}_{n_1},\mathcal{W}_{n_2})$ will be formally defined in Section~\ref{section:optimization}.

For $l_s$ bits allocated to this constellation set $\Omega_s$, the number of points of all subsets $\mathcal{W}_n$, $n = 0, ..., N - 1$, must be summed up to $2^{l_s}$, i.e., $\sum_{n=0}^{N-1} w_n = 2^{l_s}$. To find {the optimal} combination of $\{ w_0, w_1, ..., w_{N-1} \}$, an exhaustive search algorithm can be used \cite{borran2003design}; however, its computational complexity prohibits its practical implementation, especially for a higher number of bits per constellation. To overcome such a high computational complexity, we propose a constellation structure with subset $\mathcal{W}_n$, $n = 0, 1, ..., N - 1$, defined as
\begin{align}
    \mathcal{W}_n & = \{ \alpha_n \mathbf{v}_0, \alpha_n \mathbf{v}_1, ..., \alpha_n \mathbf{v}_{2^{l_v}-1}  \}, \quad n = 0, 1, ..., N-1,
    \label{eqn:multi-level_structure}
\end{align}

where $N = 2^{l_{\alpha}}$. We define the set $\Omega_{\alpha} = \{ \alpha_0, \alpha_1, \dots, \alpha_{2^{l_{\alpha}}-1} \}$ as the set of $\alpha$ that satisfies $\alpha_0<\alpha_1<\dots<\alpha_{2^{l_{\alpha}}-1}$, and we define the set $\Omega_v = \{ \mathbf{v}_0, \dots, \mathbf{v}_{2^{l_{v}}-1} \}$ as the set containing all possible unitary vectors $\mathbf{v}$, with $l_{\alpha}$ and $l_v$ be the number of bits allocated to $\Omega_{\alpha}$ and $\Omega_{v}$, respectively. 

One can observe that the set $\Omega_s$ can be seen as the Cartesian product of $\Omega_{\alpha}$ and $\Omega_{v}$, and such an observation serves two purposes: the low-complexity construction and detection of our proposed multi-level constellation design as discussed below.

\subsubsection{Low-complexity construction of the multi-level constellation} 
For our proposed structure in \eqref{eqn:multi-level_structure}, the number of possible choices for $w_n$, $n = 0, ..., N-1$, are only $l_s+1$, which corresponds to $N \in \{1,2,4,\dots,2^{l_s}\}$. In contrast, the method in \cite{borran2003design} requires an exhaustive search over approximately $e^{\pi\sqrt{2^{l_s+1}/3}}/(4*2^{l_s}\sqrt{3})$ choices of partitioning $w_n$ (Hardy–Ramanujan partition formula). Furthermore, \cite{borran2003design} requires a large number of available unitary constellations $\Omega_v$ whose number of points ranges from one to $2^{l_s}$, which requires a gigantic computational complexity. Instead, our method only requires unitary constellations whose cardinality range is the power of two, i.e., $\{1,2,4,\dots,2^{l_s}\}$, which are readily available in the literature.

\subsubsection{Low-complexity ML detection of the multi-level constellation} Instead of searching over all possible $\mathbf{s} = \alpha \mathbf{v}$, as shown in (\ref{eqn:ML_detector_splitted}), we only need to search over $\alpha$ and $\mathbf{v}$ separately, which is represented as follows:
\begin{align}
    \mathbf{v}_{\rm{ML}} & = \argmax_{\mathbf{v} \in \Omega_v} \ \rm{tr} \left( \mathbf{Y}^H \mathbf{Y} \mathbf{v}^* \mathbf{v}^T \right). \\
    \alpha_{\rm{ML}} & = \argmax_{\alpha \in \Omega_{\alpha}} \left( \frac{\alpha^2\rm{tr} \left( \mathbf{Y}^H \mathbf{Y} \mathbf{v}_{\rm{ML}}^* \mathbf{v}_{\rm{ML}}^T \right)}{\sigma^2(\sigma^2 + \alpha^2 )} - M (\sigma^2 + \alpha^2 ) \right).
    \label{eqn:ML_detector_splitted2}
\end{align}
By doing this, the ML decoding complexity of the whole constellation can be reduced from $\mathcal{O}(2^{l_{\alpha}+l_v})$ to $\mathcal{O}(2^{l_{\alpha}}) + \mathcal{O}(2^{l_v})$, where $\mathcal{O}(2^{l_{\alpha}})$ and $\mathcal{O}(2^{l_{v}})$ are the complexities of detecting $\alpha$ and $\mathbf{v}$, respectively. The decoding complexities can be further reduced if we embed special unitary constellations with low decoding complexity.

\section{Optimization of Multi-level Constellations Under Fixed Bit Allocation}
\label{section:optimization}

To find the optimal multi-level constellation $\Omega_s$, we find sub-optimal constellations $\Omega_{\alpha}$ and $\Omega_{v}$ with given  bit allocations $l_{\alpha}$ and $l_v$, respectively. Then we choose $l_{\alpha}$ and $l_v$ corresponding to constellation $\Omega_s$ with the highest KL distance. The optimization for given fixed values of $l_{\alpha}$ and $l_v$ is discussed in this Section \ref{section:optimization}.

Given the proposed structure in (\ref{eqn:multi-level_structure}), $D_{\rm{intra}} (\mathcal{W}_n)$ and $D_{\rm{inter}} (\mathcal{W}_{n_1},\mathcal{W}_{n_2})$ can be written as follows:
\begin{align}
    & D_{\rm{intra}} (\mathcal{W}_n) = \frac{\alpha_n^4}{\sigma^2(\sigma^2+\alpha_n^2)} \min_{\mathbf{v}_k \neq \mathbf{v}_i} (1 - | \mathbf{v}_k^T \mathbf{v}_i^* |^2), \\
    & D_{\rm{inter}} (\mathcal{W}_{n_1},\mathcal{W}_{n_2}) = \frac{\sigma^2 + \alpha_{n_1}^2}{\sigma^2 + \alpha_{n_2}^2} - \text{ln} \left( \frac{\sigma^2 + \alpha_{n_1} ^2}{\sigma^2 + \alpha_{n_2}^2} \right) - 1.
\end{align}
For a given $l_{\alpha}$ and $l_v$, our problem in \eqref{eq:main} can be reformulated as follows:
\begin{subequations}
    \label{eqn:optProb_optProb_multilevel}
    \begin{align}
        & \{ \Omega_{\alpha}, \Omega_v \} = \argmax_{\substack{\Omega_{\alpha}, \Omega_v \ }} \nonumber \\
        & \left( \min \left[ \min_{\mathcal{W}_n \subseteq \Omega_s } D_{\rm{intra}}(\mathcal{W}_n), \min_{\substack{n_1\neq n_2 \\ \mathcal{W}_{n_1},\mathcal{W}_{n_2} \subseteq \Omega_s }} D_{\rm{inter}}(\mathcal{W}_{n_1},\mathcal{W}_{n_2}) \right] \right), \\
        & \text{s.t.} \ \Vert \mathbf{v}_i \Vert = 1, \ \forall \mathbf{v}_i \in \Omega_v, \ \rm{card}\{\Omega_{v}\} = 2^{l_v}, \label{eqn:optProb_multilevel_constraint_v} \\
        & \qquad \frac{1}{2^{l_{\alpha}}} \sum_{\alpha_i \in \Omega_{\alpha}} \alpha_i^2 = 1, \ \rm{card}\{\Omega_{\alpha}\} = 2^{l_{\alpha}}, \label{eqn:optProb_multilevel_constraint_alpha}
    \end{align}
\end{subequations}
Since $\alpha_n^4 / (\sigma^2(\sigma^2+\alpha_n^2))$ is an increasing function of $\alpha_n$ for $\alpha_n \geq 0$, $D_{\rm{intra}} (\mathcal{W}_n)$ is minimized when $\alpha_n$ is minimized, i.e., $\alpha_n=\alpha_0$. Since $1/x-\rm{ln}(1/x)-1<x-\rm{ln}(x)-1$ for $x>1$, $D_{\rm{inter}} (\mathcal{W}_{n_1},\mathcal{W}_{n_2}) < D_{\rm{inter}} (\mathcal{W}_{n_2},\mathcal{W}_{n_1})$ for $n_1 < n_2$. Thus, the minimum inter distance must be in the cases of $D_{\rm{inter}} (\mathcal{W}_{n_1},\mathcal{W}_{n_2})$ where $n_1<n_2$. Also, since $D_{\rm{inter}}$ increases when the energy difference between two subsets increases, the minimum inter distance must be the cases of two consecutive subsets. As a result, we simplify our problem in \eqref{eqn:optProb_optProb_multilevel} as follows:
\begin{subequations}
    \begin{align}
        \{ \Omega_{\alpha}, \Omega_v \} & = \argmax_{\substack{\Omega_{\alpha}, \Omega_v \ }}
        \left\{ \min \left( D_{\rm{intra}}(\mathcal{W}_0), \ D_{\rm{inter}}(\mathcal{W}_{n},\mathcal{W}_{n+1}) \right) \right\}, \\
        \text{s.t.} \ & \eqref{eqn:optProb_multilevel_constraint_v} \: \text{and} \: \eqref{eqn:optProb_multilevel_constraint_alpha}. \nonumber
    \end{align}
\end{subequations}

We can see that the distance $D_{\rm{intra}}(\mathcal{W}_0)$ contains the variable $\mathbf{v}$ in the form of $\min_{\mathbf{v}_k \neq \mathbf{v}_i} (1 - | \mathbf{v}_k^T \mathbf{v}_i^* |^2)$, and this optimization problem of $\mathbf{v}$ is not affected by $\alpha$. Thus, we can decouple our problem into two separate optimization problems, i.e., unitary set optimization and multi-level optimization. The unitary set optimization is given by:
\begin{subequations} \label{eq:unitarySetOptimization}
    \begin{align}
    \Omega_v & = \argmax_{\Omega_v} \left\{ \min_{\mathbf{v}_k \neq \mathbf{v}_i} (1 - | \mathbf{v}_k^T \mathbf{v}_i^* |^2) \right\} \\
    \text{s.t.} \ & \eqref{eqn:optProb_multilevel_constraint_v}, \nonumber
    \end{align}
\end{subequations}
which is a classic problem called sphere packing on Grassmannian manifolds and has been well-studied in the literature  \cite{gohary2009noncoherent,zhang2011full,ngo2019cube}. Thus,  we can use any available unitary set $\Omega_v$ from the literature and focus on the multi-level optimization. Given $\Omega_v$, let $T_v= \min_{\mathbf{v}_k \neq \mathbf{v}_i} (1 - | \mathbf{v}_k^T \mathbf{v}_i^* |^2)$ be the minimum distance of the unitary constellation $\Omega_v$. The multi-level optimization problem for a given $T_v$ can be written as follows:
\begin{subequations}
    \begin{align}
        \Omega_{\alpha} & = \argmax_{\Omega_{\alpha} }
        \left\{ \min_{\substack{\alpha_0 \\ r_0,\dots,r_{N-2} }}
        \left( \frac{\alpha_0^4 T_v}{\sigma^2(\sigma^2+\alpha_0^2)}, \frac{1}{r_n} - \rm{ln}\left(\frac{1}{r_n}\right) - 1 \right) \right\}, \\
        \text{s.t.} \ & \eqref{eqn:optProb_multilevel_constraint_alpha}. \nonumber
    \end{align}
\end{subequations}
with $r_n = (\sigma^2+\alpha_{n+1}^2)/(\sigma^2+\alpha_n^2)$. The optimal $\{\bar{\alpha}_0,\bar{r}_0,\dots,\bar{r}_{N-2}\}$ must satisfy the conditions:
\begin{subequations} \label{eq:optimalSolutionOfMagnitude}
    \begin{align}
        & \bar{r}_0 = \bar{r}_1 = \dots = \bar{r}_{N-2}, \label{eq:optimalSolutionOfMagnitude_condition1} \\
        & \frac{ \bar{\alpha}_0^4 \ T_v}{\sigma^2 (\sigma^2 + \bar{\alpha}_0^2)} = \frac{1}{\bar{r}_0} - \rm{ln}\left(\frac{1}{\bar{r}_0}\right) - 1. \label{eq:optimalSolutionOfMagnitude_condition2}
    \end{align}
\end{subequations}
The proof is presented in the Appendix. Now we proceed to construct the optimal set $\Omega_{\bar{\alpha}} = \{\bar{\alpha}_0,\dots,\bar{\alpha}_{N-1}\}$ given the two conditions in (\ref{eq:optimalSolutionOfMagnitude}). From (\ref{eq:optimalSolutionOfMagnitude_condition1}), it is easy to prove that $\sigma^2 + \bar{\alpha}_i^2 = (\sigma^2 + \bar{\alpha}_0^2) (\bar{r}_0)^i$. Thus,
\begin{align}
    & 2^{l_{\alpha}}(\sigma^2 + 1) = \sum_{i=0}^{2^{l_{\alpha}}-1} (\sigma^2 + \bar{\alpha}_i^2) = (\sigma^2 + \bar{\alpha}_0^2) \sum_{i=0}^{2^{l_{\alpha}}-1} (\bar{r}_0)^i \nonumber \\
    & = (\sigma^2 + \bar{\alpha}_0^2) \left(\frac{r^{2^{l_{\alpha}}}-1}{r-1}\right). \label{eqn:optimalSolutionOfMagnitude_condition1_transformed}
\end{align}
The solution of (\ref{eq:optimalSolutionOfMagnitude_condition2}) and (\ref{eqn:optimalSolutionOfMagnitude_condition1_transformed}) can be obtained by conventional methods such as bisection or Newton method. We will present our bisection method in Algorithm \ref{alg.1}.

\begin{algorithm2e}[t]
\SetAlgoLined 
\textbf{Input:} $\sigma^2$, $l_{\alpha}$, $T_v$. \\
\textbf{Output:} ${\bar{r}_0}$, $\bar{\alpha}_0$. \\
\textbf{Parameters:} $\epsilon$, $r_{l}=1$, $r_{u}$ satisfies $\frac{1-r_{u}^{2^{l_{\alpha}}}}{1-r_{u}} \leq 2^{l_{\alpha}}\frac{1+\sigma^2}{\sigma^2}$. \\
\While{$r_{u}-r_{l} \geq \epsilon$}{
    $r = \frac{r_{u}+r_{l}}{2}$, $\alpha_0^2 = 2^{l_{\alpha}}\frac{(1+\sigma^2)(1-r)}{1-r^{2^{l_{\alpha}}}} - \sigma^2$ \\
    \eIf{$\frac{\alpha_0^4}{(\alpha_0^2+\sigma^2)\sigma^2} T_v - \left(\frac{1}{r}-\text{ln}(\frac{1}{r}) - 1\right) \leq 0$}{
        $r_{u} = r$
    }
    {
        $r_{l} = r$
    }
}
${\bar{r}_0} = r$, $\bar{\alpha}_0 = \alpha_0$.
\caption{Bisection method to solve (\ref{eq:optimalSolutionOfMagnitude_condition2}) and (\ref{eqn:optimalSolutionOfMagnitude_condition1_transformed})}
\label{alg.1}
\end{algorithm2e}

\section{Simulation Result}
\label{section:simulation_result}

In this section, we evaluate the SER performance of our proposed multi-level constellation and compare it to other constellations from the literature. Since already-optimized unitary constellations are prerequisites for our method, we evaluate our proposed multi-level constellations with two types of unitary constellations: the general unitary constellations obtained by numerically solving the optimization problem in (\ref{eq:unitarySetOptimization}) and the cube-split constellations in \cite{ngo2019cube}. While the numerically-optimized general unitary constellations have the best SER performance among all unitary constellations, the cube-split constellations have low decoding complexity while still attaining acceptable performance compared to other non-coherent schemes.

\subsection{Optimal Bit Allocation to $\Omega_{\alpha}$}
\begin{figure}[h]
    \begin{subfigure}{0.5\textwidth}
        \centering
        \includegraphics[width=9cm]{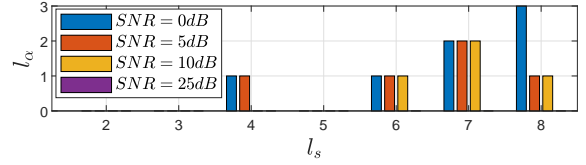}
        \caption{$K=2$, Multi-level design with cube-split constellations.}
        \label{fig:alphaBitAllocation_cubesplit_Ts2}
    \end{subfigure}
    \vfill
    \begin{subfigure}{0.5\textwidth}
        \centering
        \includegraphics[width=9cm]{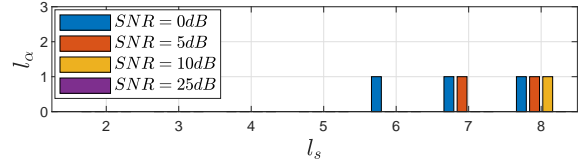}
        \caption{$K=2$, Multi-level design with general unitary constellations.}
        \label{fig:alphaBitAllocation_general_Ts2}
    \end{subfigure}
    \vfill
    \begin{subfigure}{0.5\textwidth}
        \centering
        \includegraphics[width=9cm]{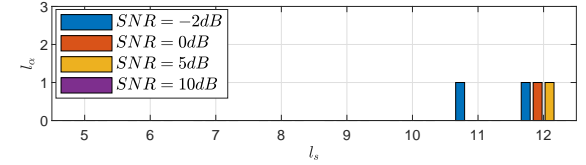}
        \caption{$K=3$, Multi-level design with general unitary constellations.}
        \label{fig:alphaBitAllocation_general_Ts3}
    \end{subfigure}
    \vfill
    \begin{subfigure}{0.5\textwidth}
        \centering
        \includegraphics[width=9cm]{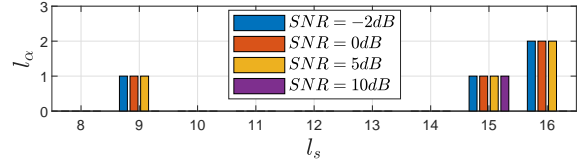}
        \caption{$K=4$, Multi-level design with cube-split constellations.}
        \label{fig:alphaBitAllocation_cubesplit_Ts4}
    \end{subfigure}
    \caption{Number of bits allocated to $\Omega_{\alpha}$ for optimal multi-level design.}
    \label{fig:bitAllocationPlot}
\end{figure}
First, we plot the number of bits allocated to $\Omega_{\alpha}$. Since our multi-level design is SNR-adaptive, given a collection of unitary constellations $\Omega_v$, the number of bits allocated to $\Omega_{\alpha}$,  and hence, to $\Omega_v$, will only depend on the SNR level. In general, if the number of bits allocated to $\Omega_{\alpha}$ is greater than zero, it means our multi-level design achieved higher KL distance than unitary constellations. In contrast, if the number of bits allocated to $\Omega_{\alpha}$ is equal to zero, our multi-level design simply converges to unitary constellations, and no performance gain is achieved from our multi-level design. Figure \ref{fig:bitAllocationPlot} shows that the number of bits allocated to $\Omega_{\alpha}$ tends to decrease when the SNR increases and converges to 1-level design (unitary constellations) in the high SNR regime. This is in line with the statement of optimality of unitary constellations in the high SNR regime in \cite{borran2003design}. It can also be seen from Fig. \ref{fig:bitAllocationPlot} that for higher $K$, our multi-level design still has advantages over unitary constellations, though less frequently than the case of $K=2$.

\subsection{SER Performance Evaluation}
\begin{figure}
    \centering
    \includegraphics[width=8cm]{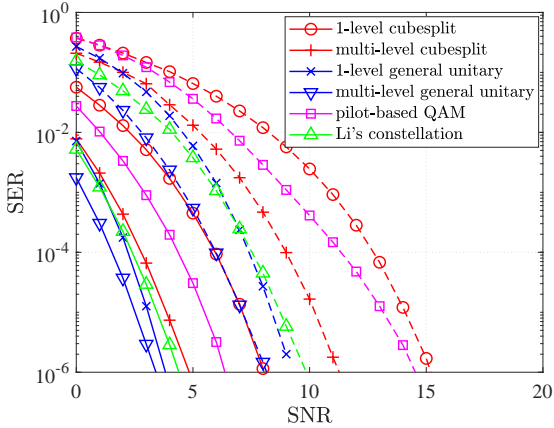}
    \caption{SER comparison between the proposed multi-level design, 1-level design, and other non-coherent schemes in case of $K=2$ and $M=256$. Solid line: $l_s=6$. Dashed line: $l_s=8$}
    \label{fig:SER_multilevel_vs_1level_Ts2}
\end{figure}

In this sub-section, we evaluate the SER performance gain achieved by our multi-level design (combined with unitary constellations) when compared to only unitary constellations without multi-level design. We also use two non-coherent schemes as baselines for comparisons: non-coherent detection of pilot-based QAM, which includes a pilot symbol and a sequence of QAM symbols, and the multi-level constellations from \cite{li2021constellation}. Fig. \ref{fig:SER_multilevel_vs_1level_Ts2} shows that multi-level design improves the SER performance in both general unitary constellations and cube-split constellations for both low and high SNR regimes. By embedding our multi-level design into cube-split constellations, the cube-split design achieves a significant improvement in SER performance and outperforms the conventional pilot-based QAM scheme. Additionally, the proposed multi-level design improves the performance of general unitary constellations and also outperforms both of the reference schemes. Please note that the multi-level design will converge to 1-level design when SNR further increases, as discussed in Fig. \ref{fig:bitAllocationPlot}.

\begin{figure}
    \centering
    \includegraphics[width=8cm]{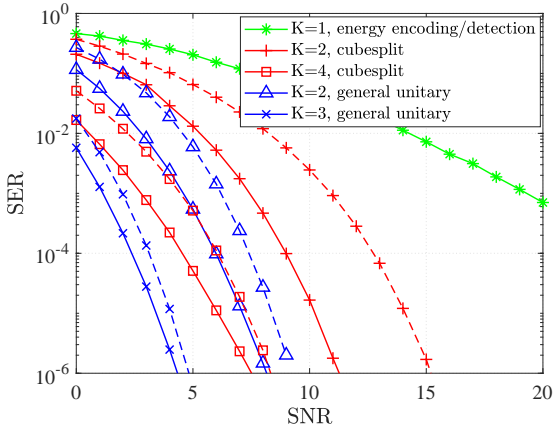}
    \caption{SER comparison between the proposed multi-level design, 1-level design, and other non-coherent schemes for different values of $K$ and $M=256$.}
    \label{fig:SER_multilevel_vs_1level_differentK}
\end{figure}
Since our method can be applied to any number of time slots $K$, we also evaluate the SER performance among different values of $K$ given the same SE of 4 bits/Hz/s. In the case of $K=1$, energy encoding and detection in \cite{han2022constellation}, which is also a specific case of our multi-level design for $K=1$, will be used as a reference. Firstly, Fig. \ref{fig:SER_multilevel_vs_1level_differentK} shows that our multi-level design also improves the performance of the cube-split constellation for $K=4$ and general unitary constellation for $K=3$. In general, since the overall performance of non-coherent constellations improves when $K$ increases, it is reasonable to apply our multi-level design to enhance the performance of unitary constellations in a higher number of time slots. In contrast, the multi-level designs in \cite{li2021constellation} (which is only designed for $K=2$) and \cite{borran2003design} (which has high complexity and therefore cannot be applied for large $K$) cannot achieve performance gain by increasing the number of time slots.

\section{Conclusion}
\label{section:conclusion}
In this paper, we proposed a low-complexity design for multi-level constellations based on KL divergence. Our multi-level design is shown to improve the SER performance of non-coherent schemes based on unitary constellations as well as outperforms other non-coherent schemes while having low complexity in construction and detection.

\begin{appendix}

\label{appendix_A}
We prove the set $\Omega_{\bar{\alpha}}$ is the optimal solution of (\ref{eq:optimalSolutionOfMagnitude}) by contradiction. In this case, the objective value corresponding to $\Omega_{\bar{\alpha}}$ is $\frac{ \bar{\alpha}_0^4 \ T(k_{\alpha})}{\sigma^2 (\sigma^2 + \bar{\alpha}_0^2)}$ or $\frac{1}{\bar{r}_0} - \rm{ln}\left(\frac{1}{\bar{r}_0}\right) - 1$. Now, assume that $\Omega_{\bar{\alpha}}$ is not the optimal solution, then there exists at least one other set $\Omega_{\hat{\alpha}} = \{\hat{\alpha}_0,...,\hat{\alpha}_{N-1} \}$ and its corresponding $\{\hat{r}_0,\dots,\hat{r}_{N-2}\}$ that has higher objective value. This  is equivalent to:
\begin{align}
    & \frac{ \hat{\alpha}_0^4 \ T_v}{\sigma^2 (\sigma^2 + \hat{\alpha}_0^2)} >  \frac{ \bar{\alpha}_0^4 \ T_v}{\sigma^2 (\sigma^2 + \bar{\alpha}_0^2)} \label{eq:first_inequality} . \\
    & \frac{1}{\hat{r}_n} - \rm{ln}\left(\frac{1}{\hat{r}_n}\right) - 1 > \frac{1}{\bar{r}_n} - \rm{ln}\left(\frac{1}{\bar{r}_n}\right) - 1, \ \forall n \in \{0,\dots,N-2\} \label{eq:second_inequality} .
\end{align}
Since $\frac{ \alpha_0^4 }{\sigma^2 (\sigma^2 + \alpha_0^2)}$ is an increasing function of $\alpha_0 > 0$ and $1/r - \rm{ln}(1/r) - 1$ is an increasing function of $r>1$, (\ref{eq:first_inequality}) and (\ref{eq:second_inequality}) are equivalent to $\hat{\alpha}_0 > \bar{\alpha_0}$ and $\hat{r}_n > \hat{r}_n$, respectively. Thus,
\begin{align}
    \sigma^2 + \hat{\alpha}_n^2 & =  (\sigma^2 + \hat{\alpha}_0^2) \prod_{i=0}^{n-1} \hat{r}_i  > (\sigma^2 + \bar{\alpha}_0^2) \prod_{i=0}^{n-1} \bar{r}_i = \sigma^2 +  \bar{\alpha}_n^2
\end{align}
$\Omega_{\hat{\alpha}}$ clearly does not satisfy the power constraints because:
\begin{equation}
    \frac{1}{2^{k_{\alpha}}} \sum_{i=0}^{2^{k_{\alpha}-1}} \hat{\alpha}_i^2 > \frac{1}{2^{k_{\alpha}}} \sum_{i=0}^{2^{k_{\alpha}-1}} \bar{\alpha}_i^2 = 1,
\end{equation}
which concludes the proof. $\hfill \blacksquare$

\end{appendix}

\bibliographystyle{IEEEtran}
\bibliography{IEEEabrv, references}

\end{document}